\documentclass[conference]{IEEEtran}

\usepackage{graphicx}[pdf]
\usepackage{subcaption}
\usepackage{amsmath}
\usepackage{cite}
\usepackage{algorithm}
\usepackage{algpseudocode}

\makeatletter
\newenvironment{smartcontract}[1][htb]{%
    \renewcommand{\ALG@name}{Smart Contract}
   \begin{algorithm}[#1]%
  }{\end{algorithm}}
\makeatother

\algblockdefx{Contract}{EndContract}[1]{\textbf{contract} #1}{\textbf{end contract}}
\algblockdefx{Action}{EndAction}[1]{\textbf{action} #1}{\textbf{end action}}
\algblockdefx{Record}{EndRecord}[1]{\textbf{record} #1}{\textbf{end record}}

\begin{document}

\title{Blockchain-based Trust Information Storage in Crowdsourced IoT Services}

\author{\IEEEauthorblockN{Mohammed Bahutair}
\IEEEauthorblockA{University of Sydney\\
Sydney, Australia\\
mbah6158@uni.sydney.edu.au}
\and
\IEEEauthorblockN{Athman Bouguettaya}
\IEEEauthorblockA{University of Sydney\\
Sydney, Australia\\
athman.bouguettaya@sydney.edu.au}
}

\maketitle

\begin{abstract}
We propose a novel distributed integrity-preserving framework for storing \emph{trust information} in crowdsourced IoT environments. The \emph{integrity} and \emph{availability} of the trust information is paramount to ensure accurate trust assessment. Our proposed framework leverages the \emph{blockchain} to build a distributed storage medium for trust-related information that ensures its integrity. We propose a \emph{geo-scoping} approach, which ensures that trust-related information is only available where needed, thus, enabling fast access and storage space preservation. We conduct several experiments using real datasets to highlight the effectiveness of our framework.
\end{abstract}


\section{Introduction}
\label{section:introduction}

The proliferation of Internet-based platforms and mobile applications has given rise to the concept of the \emph{shared economy}. It has introduced a fundamental shift on the way people go about their social and economic activities \cite{huckle2016internet}. Traditionally, there is a clear distinction between service providers and consumers. Conversely in a shared economy, any \emph{thing} can be a provider, consumer, or both at the same time. A fundamental prerequisite for a successful deployment of the shared economy is the establishment of \emph{trust} between its entities.


The emergence of \emph{Internet of Things} (IoT) has opened opportunities for a \emph{digital shared economy}. In essence, the Internet of Things (or IoT) is an ecosystem where \emph{things} (e.g., shoes, cars, and watches) are interconnected to share information through the Internet \cite{gubbi2013internet}. IoT has paved the way for a multitude of applications such as smart cities and smart homes \cite{gubbi2013internet}. \emph{Crowdsourcing} is a fertile ground for leveraging IoT to provide and consume \emph{services}. For example, a smartphone (\emph{service provider}) may elect to offer its computing resources (e.g., CPU and memory) to a nearby smartwatch (\emph{service consumer}). The smartwatch (which has limited computational power) may use the offered service to delegate some of its computationally-intensive tasks to the smartphone. IoT devices can offer a variety of service types. Examples of such services include but not limited to: \emph{compute resource} \cite{habak2015femto}, \emph{energy sharing} \cite{lakhdari2018crowdsourcing, dhungana2019exploiting, bulut2018crowdcharging}, \emph{environmental sensing} \cite{kelly2013towards}, and \emph{WiFi hotspot} \cite{neiat2017crowdsourced} services. IoT services generally consists of two parts: \emph{functional} and \emph{non-functional} parts \cite{neiat2017crowdsourced}. The functional part describes the \emph{purpose} of the service, whereas the non-functional part represents the qualities that surround the delivery of the functionalities.



\begin{figure*}
    \centering
    \includegraphics[width=0.85\textwidth]{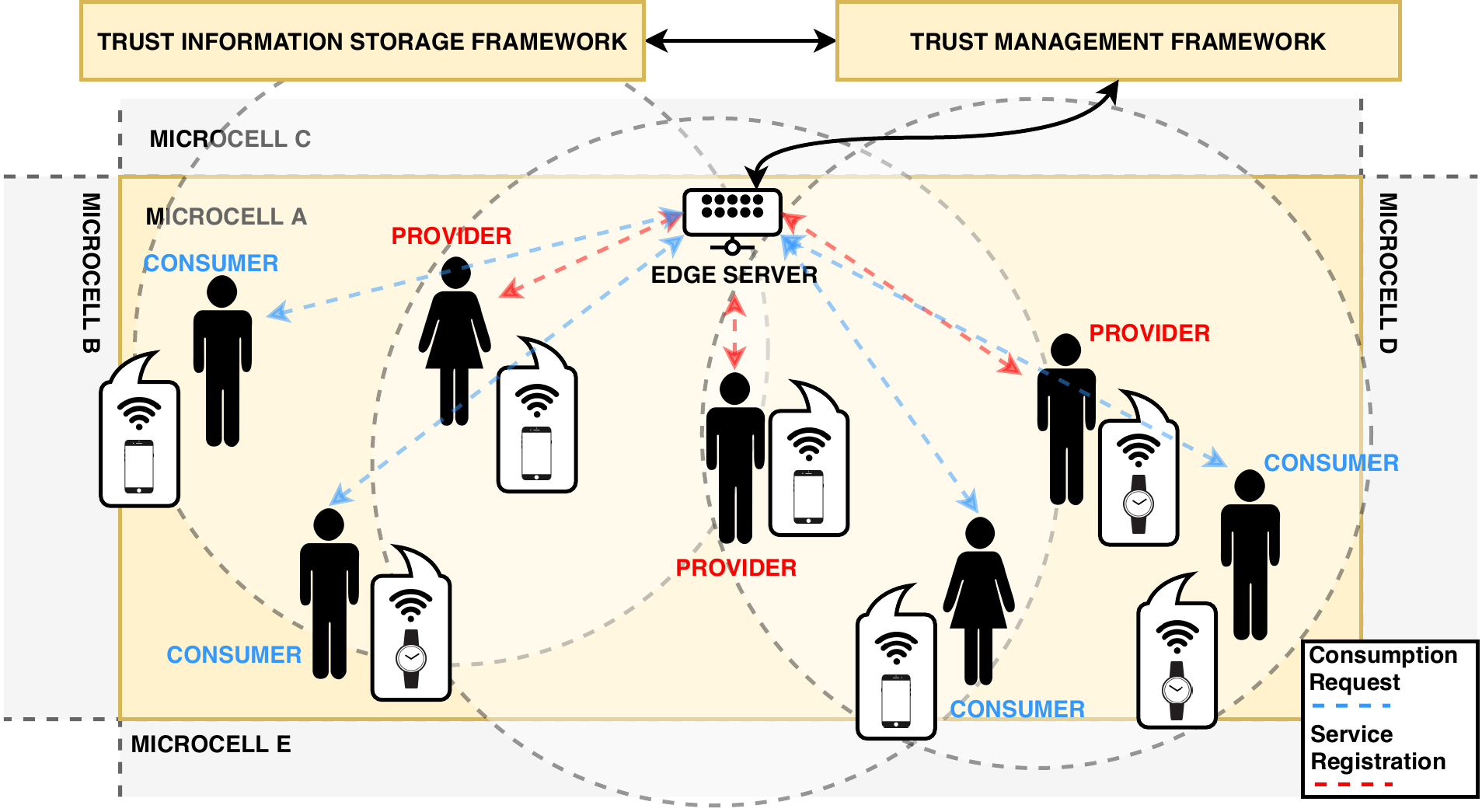}
    \caption{The general architecture of an IoT service crowdsourcing environment.}
    \label{fig:architecture}
\end{figure*}

Crowdsourced IoT services can offer potential benefits in terms of \emph{convenience} and \emph{resource utilization} \cite{habak2015femto}. However, several challenges may arise in such environments. One key challenge is the \emph{establishment of trust} among IoT service providers and consumers \cite{bahutair2019adaptive}. For example, assume a WiFi hotspot crowdsourcing environment where IoT devices shared their Internet with other IoT devices \cite{neiat2017crowdsourced}. On one hand, a service consumer may require some assurances that their data is not disclosed. On the other hand, a service provider may require that their services are not being used for illegal actions. In other words, \emph{mutual trust} should be established between the provider and consumer for a successful service provisioning. Existing approaches (e.g., \cite{bahutair2019adaptive} and \cite{bahutair2021multiperspective}), proposed frameworks and techniques for assessing the trustworthiness of IoT services prior to service consumption. The frameworks account for the \emph{dynamic nature} typically exhibited by IoT environments. We relied mainly on the \emph{inherent characteristics} of IoT services as well as \emph{context parameters} to perform the trust assessment. Such parameters include device reputation, device's model, device's operating system, owner's rating, etc. We refer to these parameters and any data used for trust evaluation as \emph{trust information}. Ensuring the \emph{veracity} of the trust information is crucial since inaccurate trust information may potentially lead to unreliable trust evaluation.

One of the main requirements to ensure the correctness of trust information is a \emph{storage medium} where data is guaranteed to be protected from tampering. In other words, it is paramount to have a \emph{trust information storage framework} that manages the data used for trust assessment and preserves its integrity against unauthorized alternations. The \emph{blockchain} \cite{nakamoto2019bitcoin} is a prime candidate to serve as the basis for such a framework. The blockchain is a \emph{distributed} \emph{integrity-preserving} storage platform. The data in the blockchain is organized into \emph{blocks}. Blocks are connected into one single \emph{chain} (hence the name blockchain). In other words, each block has a reference to the block before it. The blockchain infrastructure consists of a network of computers typically referred to as \emph{nodes}. One of the main goals in blockchain networks is for its nodes to reach \emph{consensus}. Consensus is reached when all nodes in the network have the same copy of the blockchain. Several methods (i.e., \emph{consensus mechanisms}) have been proposed that guarantee consensus between nodes (e.g., Proof-of-Work \cite{nakamoto2019bitcoin}, Proof-of-Stake \cite{king2012ppcoin}, and The Stellar Consensus Protocol \cite{lokhava2019fast, mazieres2015stellar}). The Proof-of-Work mechanism is widely used, however, a node has to be computationally capable to participate. Essentially, a node's role in such networks is to \emph{mine} new blocks. Mining is the process of computing a unique value (called a \emph{hash}) for a given block that satisfies a certain condition. Computing the hash is carried out in a brute-force fashion. The goal is to ensure that \emph{miners} have spent the time to acquire the hash. Proof-of-Work (while it can guarantee data integrity) is not applicable for IoT environments, since IoT devices generally have low processing power. A more suitable alternative is the Stellar Consensus Protocol, which relies on \emph{federated Byzantine agreement}. Nodes in the protocol can reach consensus by exchanging messages as opposed to performing computationally extensive tasks, which makes it a prime candidate for IoT environments.

We identify three main requirements for such a trust information storage framework: \emph{integrity}, \emph{accessibility}, and \emph{availability}. \emph{Integrity} indicates that the trust information remains intact with no unauthorized alterations. For example, a trust management framework (e.g., \cite{bahutair2021multiperspective}), may request a device's reputation to assess a given service's trust. The storage framework, which the trust framework is requesting the information from, should guarantee that the provided information is correct and has not been tampered with. Failing to achieve this requirement would result in an inaccurate (and potentially unreliable) trust assessment. \emph{Accessibility} indicates that the storage framework should provide a way to address the stored trust information. The complexity of how trust information is accessed depends mainly on the storage framework's type in terms of deployment. A storage framework can either be \emph{central} or \emph{distributed}. The distributed option is more suited for IoT environments since a central deployment may be impractical due to a large number of devices. However, a distributed storage framework poses data-access-related challenges. For example, assume a trust management framework requests certain data from the storage framework. The storage framework would need first to \emph{locate} where the data is stored. One way to mitigate such a challenge is by duplicating the entire trust information collection at each data storage location. However, this may result in a waste of storage space since not all information is used equally at every location. Therefore, the trust information should be distributed based on their \emph{usage}. Finally, the \emph{availability} requirement entails that the distribution of the trust information should satisfy any potential request.

The contribution of the paper focuses on a \emph{trust information storage framework} that satisfies the requirements stated earlier. Specifically, we propose:
\begin{itemize}
    \item A novel \emph{blockchain-based} framework that fits the dynamic nature of IoT environments. The framework aims at preserving the integrity of trust information at all times. The framework leverages the Stellar Consensus Protocol \cite{lokhava2019fast, mazieres2015stellar}; a light weight consensus mechanism, which is suitable for low-power devices such as found in IoT environments.

    \item A set of techniques that utilize Smart Contracts \cite{buterin2014next} to structure the trust information, thus ensuring the accessibility of the information efficiently. 

    \item To implement a set of methods that analyzes the stored trust information and ensures their availability for potential requests.
\end{itemize}

\begin{figure*}
    \centering
    \includegraphics[width=0.85\textwidth]{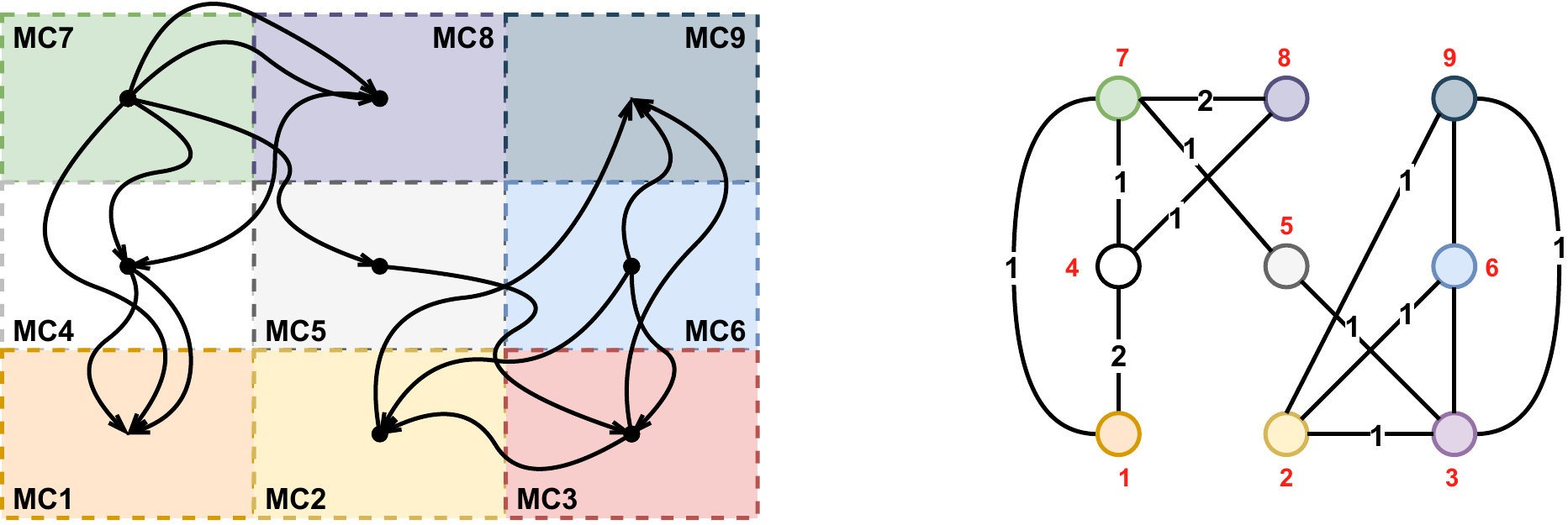}
    \caption{Converting a set of microcells into its equivalent graph using movement patterns.}
    \label{fig:microcell_to_graph}
\end{figure*}

\subsection{Motivation Scenario}
\label{section:motivation_scenario}
We use the following motivation scenario to highlight the importance of our work. Assume an IoT environment where users leverage their IoT devices to offer/consume WiFi hotspot services \cite{neiat2017crowdsourced} among each other. Suppose user $A$ uses their smartwatch to \emph{provide} their WiFi hotspot service to nearby IoT devices (i.e., service provider). Conversely, assume a nearby user $B$ that searches for a WiFi hotspot service to consume using their smartwatch (i.e., service consumer). Consumer $B$ finds out that provider $A$ is in their vicinity and can consume their service. However, consumer $B$ wishes to ensure that provider $A$ is trustworthy before service consumption. Therefore, the consumer needs to assess the service provider’s trust level using a \emph{trust management framework} (e.g., \cite{bahutair2021multiperspective}). The trust management framework relies on \emph{trust information} related to service provider $A$ to evaluate their trust. Such trust information has to be stored in an \emph{integrity-preserving} storage solution, since any unauthorized alternation would eventually result in inaccurate trust assessment.


\section{Preliminaries}
\label{section:preliminaries_and_problem_definition}
We assume an IoT crowdsourcing framework where IoT devices provide and consume services to and from other IoT devices \cite{lakhdari2020composing} (See Fig. \ref{fig:architecture}). In such a framework, we assume that IoT devices are \emph{spatially} grouped into \emph{microcells} (e.g., in Fig. \ref{fig:architecture} Microcells $A$ through $E$). IoT devices that lie in the same microcell can provide/consume services among each other. Service provisioning in the framework occurs as follows. An IoT service provider $p_S$ announces that it would start providing a service $S$ to other devices in the same microcell using their IoT device $d$. We use the term \emph{service session} $S_{session}$ to refer to the period where a service is currently available for consumption by other consumers. An IoT consumer $c_S$ would look for available service sessions in their microcell. The consumer would rely on a trust management framework (e.g., \cite{bahutair2019adaptive}, \cite{bahutair2021multiperspective}) to assess IoT services' trustworthiness prior to service consumption. The trust management framework leverages \emph{trust information} to evaluate a service's trustworthiness. Examples of trust information could be the rating of the owner, device type, number of current consumers, etc. Trust information is stored at a \emph{distributed} trust information storage framework. The trust management framework requests the trust information it needs from the storage framework. Upon trust evaluation, the consumer would decide whether to use the service based on the trust assessment. A rating (i.e., consumer-sourced trust information) might be given to the service by the consumer after service consumption ends. Ratings and other service properties (i.e., trust information) are then stored into the trust information storage framework.

\section{Blockchain-Based Trust Information Storage}
\label{section:framework}

We propose a blockchain-based framework for storing \emph{trust information}. The framework covers three main aspects: (1) \emph{infrastructure}, which covers how the proposed framework interacts with existing microcells, (2) \emph{trust information block structure}, which addresses how trust information is stored, accessed, and structured, and (3) \emph{block mining}, which covers the mining aspect of the blockchain taken into consideration the limitations of IoT environments.

\subsection{Infrastructure}
%

The goal of the proposed framework is to store trust information in a distributed fashion. Trust information includes all data needed by a trust management framework to evaluate IoT services. Such data represents previous service sessions information and service providers. In other words, any piece of trust information is generated because a service session has occurred. Recall that a service session starts at a specific spatial microcell. Note also that services are provided by IoT devices typically carried/worn by people. As a result, any data generated at a specific microcell, might not be needed in another geographically distant microcell. Hence, some data is not required to be available at some microcells since they are not needed at said microcells. Therefore, we introduce the concept of \emph{trust information scoping}, where a particular piece of trust information is \emph{scoped} to a single microcell (or a group of microcells) where the data is needed. In other words, for each group of microcells that share the same trust information, a \emph{scope} is created to hold information relevant to them. This drastically reduces storage use, since only important data is preserved. 

A trust information scope is created by first identifying \emph{microcells that share the same trust information}. We refer to such microcells as \emph{scoped microcells}. Scoped microcells are detected by relying on the movements of service providers between microcells. Recall that trust information is generated based on service providers (e.g., rating, device model, availability, etc). Therefore, microcells, where the same providers offer their services, end up sharing the same trust information. Hence, such microcells should be grouped together and share the same trust information scope. The rationale behind relying on movement patterns is that people tend to have a specific routine when it comes to their daily movements. For example, an employee might have a daily habit where they go to a coffee shop then their workplace then a restaurant for dinner. All these places can potentially be microcells, where IoT devices locating in them may share services among them. For example, the employee earlier may be a service provider that chooses to offer their service at the coffee place and restaurant. Therefore, the two microcells (coffee place and restaurant) can be considered as scoped microcells since they may share trust information regarding the employee (i.e., service provider).

An IoT service provider $p_\mathcal{S}$ at a microcell communicates first with the microcell edge server whenever they decide to start a service session $\mathcal{S}_{session}$ (i.e., a state where they are ready to serve consumers). The service provider shares whether it had offered its service previously with the microcell. In case it had, it also shares the $ID$ of the previous microcell. This way, microcells have the necessary data to identify the movements of their service providers. 

Our aim at the end is to divide the set of microcells into a set of scoped microcells. By monitoring the movements of service providers, a given microcell ends up having a list of microcells that share the same set of service providers. We use this information to convert the microcells into a graph as shown in Fig. \ref{fig:microcell_to_graph}. The figure depicts a set of nine microcells (on the left) and their equivalent graph representation (on the right). Each path between two microcells represents a service provider movement. For example, $MC5$ and $MC7$ share one single service provider. Microcells $MC1$ and $MC4$ share two service providers since there are two paths going from $MC4$ to $MC1$. We use the set of microcells and paths between them to generate a new weighted graph. Each vertex in the graph represents a microcell. Edges on the graph represent the paths between microcells. For instance, the path between $MC5$ and $MC7$ represents the edge between their respective vertices on the graph. The weight of the edge is governed by the number of paths between two given microcells. For example, the edge between vertices 1 and 4 has a weight of 2 since there are two paths between their respective microcells.

We use graph theory to detect scoped microcells using the generated graph. More precisely, we apply community detection on the microcells graph. Community detection techniques detect vertices in a graph that have some similarity between them. Since our graph is generated based on the shared information between microcells, each detected community would include vertices (aka microcells) that share similar trust information (essentially scoped microcells). One crucial point regarding detecting communities in our environment is that it has to be carried out in a decentralized fashion. Our environment is assumed to be fully distributed with no central points to manage it. Hence, microcells should be able to detect their scope (aka community) without relying on a central authority. We use the technique proposed in \cite{raghavan2007near} to detect the communities in our graph. The technique starts by labeling every vertex in the graph with a random label. Then, every vertex looks to its neighboring vertices' labels. The vertex then labels itself with the one with the highest majority. When two labels have the highest majority, a random choice is made. This can reduce the performance of the technique. Therefore, we rely on an optimized version proposed in \cite{lakhdari2016link} that picks the label based on edge strength rather than making a random selection.

A blockchain network is created for every group of scoped microcells. The blockchain holds the information relevant to the microcells in the scope. The nodes of the network consist of the edge servers of the microcells as well as the IoT devices inside the microcells. IoT devices are leveraged for mining new blocks (discussed in detail in Section \ref{section:mining}). Note that, the technique should accommodate for cases where the structure of the graph changes (e.g., movement pattern changes). However, in this work, we assume that such changes rarely occur. Dynamic movement patterns will be investigated in future work.

\begin{figure}
    \centering
    \includegraphics[width=0.35\textwidth]{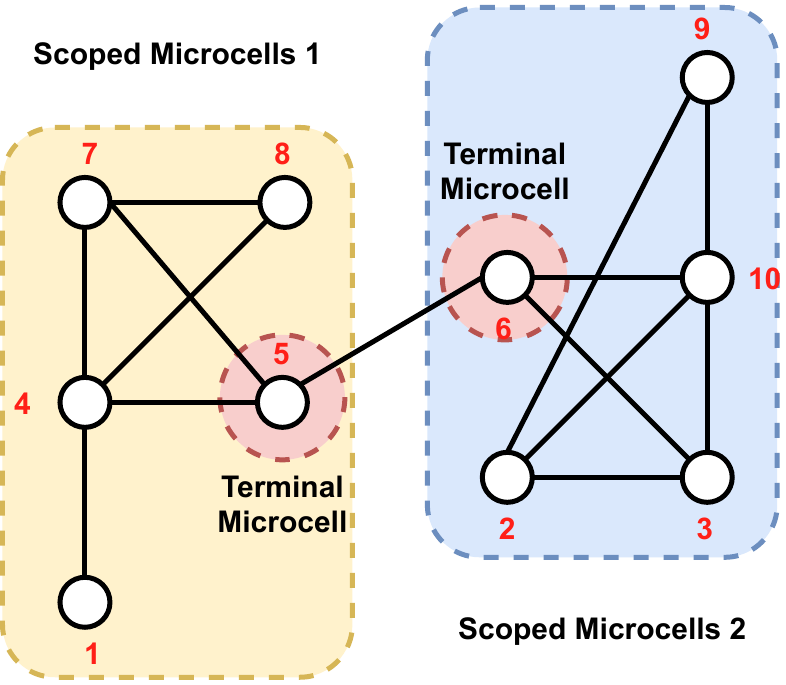}
    \caption{Terminal microcells example.}
    \label{fig:terminal_microcells}
\end{figure}

The aim of scoping trust information is to ensure that only required information is stored, thus preserving storage space. However, on rare occasions, a microcell might require trust information that only exists in an outer blockchain network (or scope). To overcome such scenarios, each scoped microcells promote one of their microcells to be an access point to other blockchain networks for the scope. We refer to such microcells as \emph{terminal microcells}. Terminal microcells are selected based on the number of edges with outer blockchain networks. For example, Fig. \ref{fig:terminal_microcells} depicts two terminal microcells for two groups of scoped microcells. Each group of scoped microcells promoted a microcell that has a connection to the other group. For instance, the first group selected microcell 5 to be the terminal microcell since it is the only one that has a connection to the other group. 

\subsection{Smart Contract-based Trust Information Handling}
A smart contract \cite{buterin2014next} is a small program that is stored in the blockchain. A major advantage of smart contracts is their \emph{immutability}. In that respect, once a smart contract is deployed, it cannot be modified thus preventing any unauthorized alteration. We leverage smart contracts to store and retrieve trust information into the blockchain. The smart contract is deployed at each available blockchain network (i.e., for each scoped microcells group). The addresses of the smart contracts are stored in their respective edge servers. When a consumer in a given microcell decides to consume an IoT service, it first tries to evaluate the service's trustworthiness. A trust management framework communicates with the edge server to obtain the address of the smart contract. Then, necessary trust information is fetched from the smart contract. By the end of service consumption, a consumer might wish to add new trust information based on their experience. The consumer would request the address of the smart contract from the microcell's edge server. Finally, the consumer would submit their trust information to the smart contract, which will eventually store it on the blockchain.

Our proposed smart contract, namely \texttt{TrustInformationHandler} consists of three main parts: (1) an \texttt{information} cache for storing the trust information, (2) a \texttt{store} action to handle saving the data into the blockchain, and (3) a \texttt{retrieve} action to get trust information from the blockchain. The \texttt{information} cache is a hash map that uses a key to address the information. The key can be one aspect of an IoT service, e.g., service owner or IoT device. The actual type of the information is a record, which in turn consists of two fields: (1) a \texttt{timestamp} field to store how recent the information is, and (2) a \texttt{data} field that stores the actual data (e.g., the rating of the owner). The action \texttt{store} is typically invoked by a service consumer to store any trust information they might wish to report regarding their IoT service. The consumer should pass in the type of trust information (e.g., the service owner rating) as the key, and the actual data (e.g., the value of their rating). The action \texttt{retrieve} is generally used by a trust management framework whenever a trust assessment is required. The framework would pass the trust information it needs, to which the smart contract would return the actual information. The implementation of the smart contract is listed in Smart Contract \ref{algorithm:smart_contract}.

\begin{smartcontract}
    \caption{Trust Information Handler}
    \label{algorithm:smart_contract}
    \begin{algorithmic}[1]
        \Contract{TrustInformationHandler}
            \Record{TrustInformation}
            \State timestamp
            \State data
            \EndRecord
            
            \State
            
            \State information $\leftarrow$ \{\}
            
            \State
            
            \Action{store(key, data)}
                \State record $\leftarrow$ TrustInformation(
                \State\space\space\space timestamp $\leftarrow$ now
                \State\space\space\space data $\leftarrow$ data
                \State )
                \State information[key] $\leftarrow$ record
            \EndAction
            
            \State
            
            \Action{retrieve(key)}
                \State value $\leftarrow$ information[key]
                \State \Return value
            \EndAction{}
            
        \EndContract
    \end{algorithmic}
\end{smartcontract}

\subsection{Mining}
\label{section:mining}
Blockchain stores \emph{blocks} of data in a distributed fashion across the nodes in its network. A major challenge in such a case is ensuring that the data between nodes is in sync and preserving the integrity of the data. \emph{Consensus} in blockchain refers to a state where all nodes in a blockchain network agree on the data they hold. Data integrity on the blockchain is achieved by ensuring that nodes are in the consensus state. In such a state, if a malicious node tries to manipulate the data, the network would be able to detect and prevent it. A \emph{consensus mechanism} is a protocol that nodes in the network follow to reach the consensus state. The \emph{Proof-of-Work} consensus mechanism \cite{nakamoto2019bitcoin} is by far the most widely used. It works by asking nodes to \emph{compete} in solving a mathematical puzzle whenever a new block is to be generated. Essentially, every node would have to solve the puzzle whenever it wishes to write data on the blockchain Such puzzles have to be solved in a short period of time. The node that succeeds in solving the puzzle first gets to add its data to the network. Therefore, having a computationally capable node is necessary for generating blocks.

We propose to leverage the large number of devices to create new blocks. Such blocks would eventually hold the trust information. While Proof-of-Work guarantees data integrity, it is not suitable for IoT environments. IoT devices have typically low processing power. Conversely, Proof-of-Work is a computationally-intensive mechanism. One way to adapt the Proof-of-Work to the less-capable IoT devices is to reduce the \emph{difficulty} of the mechanism. More specifically, we reduce the complexity of the mathematical puzzles so that devices with lesser processing power would be able to solve them within the required time period. In such a way, the required processing power by the mechanism would match with that of IoT devices. However, lessening the difficulty of a blockchain network has a critical disadvantage. The fact that IoT devices would generally contribute to the network does not limit the access to the network from other more capable devices. In other words, a highly capable malicious node can join the network and easily solve the relatively easy puzzle faster than any IoT device. 

The \emph{Stellar Consensus Protocol (SCP)} \cite{lokhava2019fast, mazieres2015stellar} is another consensus mechanism that does not rely on the nodes' (essentially IoT devices) computational power. The mechanism relies on \emph{Federated Byzantine Agreement (FBA)} \cite{mazieres2015stellar} to achieve consensus. FBA is a distributed form of the traditional Byzantine Agreement (BA) \cite{pease1980reaching, lamport2019byzantine}. BA guarantees data integrity across a network of nodes given that $N = 3f + 1$, where $N$ is the number of nodes and $f$ is the maximum number of malicious nodes the system can tolerate. For example, a system with $N = 10$ nodes of which $f = 3$ are malicious would always guarantee data integrity. Typically, a BA system would have a centralized point that manages the nodes; i.e., node additions and removals. Such a system is not suitable for environments such as the IoT. IoT devices can come and go, therefore, a centralized point that manages them would be impractical. The FBA's purpose is to implement BA systems without the need for a centralized point to manage nodes. Essentially, any node can join and leave without disrupting the network.

FBA introduces the concept of \emph{quorum slices}. Each node $v$ in the network may have one or more quorum slices. Each quorum slice contains a set of nodes $\mathcal{S}$ that $v$ trusts. A \emph{quorum} is another concept that refers to the set of nodes, where each node has at least a single quorum slice in the quorum. For instance, assume the nodes $A$, $B$, and $C$. Suppose that $A$ trusts $B$ and $C$, $B$ trusts $C$, and $C$ trusts $A$ and $B$. The node $A$'s quorum slice is ${B, C}$, $B$'s quorum slice is ${C}$, and $C$'s quorum slice is ${A, B}$.  Nodes $A$, $B$, and $C$ can together form a valid quorum since each node has its own slice in the quorum. However, if we introduce a node $D$ into $A$'s quorum slice, the three nodes $A$, $B$, and $C$ can no longer form a quorum since $A$'s slice is not entirely in the quorum. It is worth noting that FBA networks (such as the Stellar Network) could potentially have several quorums. As stated earlier, for a quorum to be valid, it has to include at least one quorum slice for all of its nodes. Additionally, a quorum should also intersect with every other quorum in the network with at least one node. Nodes in FBA reach consensus by exchanging messages regarding a piece of data to be written on the blockchain. FBA states that consensus is reached if at least one quorum agrees on the data. A quorum agreement involves that all nodes in the quorum accept the data as valid data.

We opt to utilize the Stellar Consensus Protocol in our IoT environment. Nodes in the network include IoT devices and microcells' edge servers. The dynamic nature of IoT devices should not affect the efficiency of the blockchain since FBA adapts for frequent removals and additions of IoT devices. Typically, IoT devices along with edge servers would be used to confirm that a single piece of data is valid (the mining part). However, edge server nodes will be responsible to preserve the blockchain data.

\begin{figure*}[!t]
\centering
\begin{minipage}{.42\textwidth}
    \centering
    \includegraphics[width=1.0\textwidth]{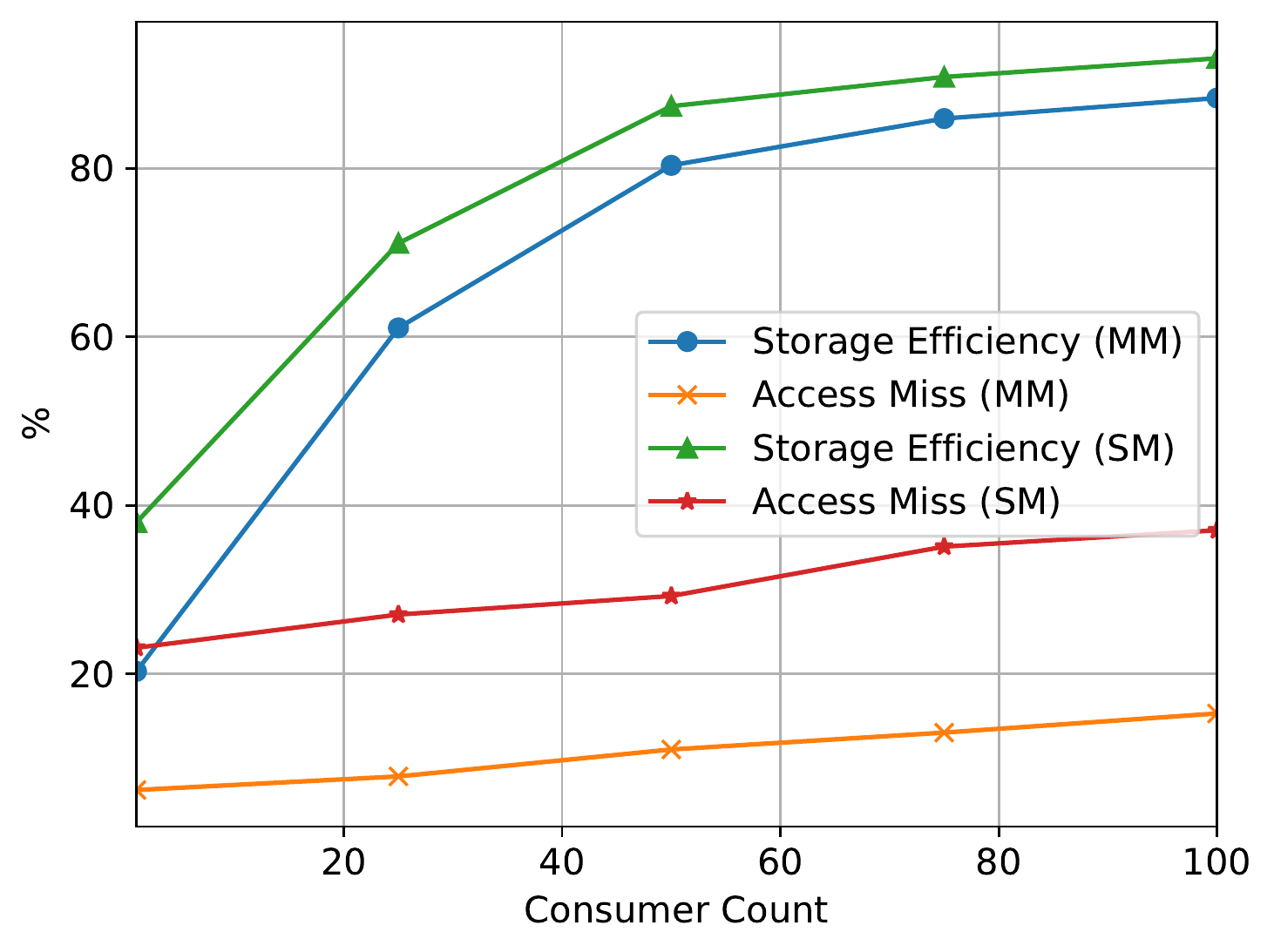}
    \caption{The storage efficiency and access misses of the framework when varying the number of consumers (Gowalla).}
    \label{figure:consumer_count_gowalla}
\end{minipage}%
\hspace{0.5cm}
\begin{minipage}{.42\textwidth}
    \centering
    \includegraphics[width=1.0\textwidth]{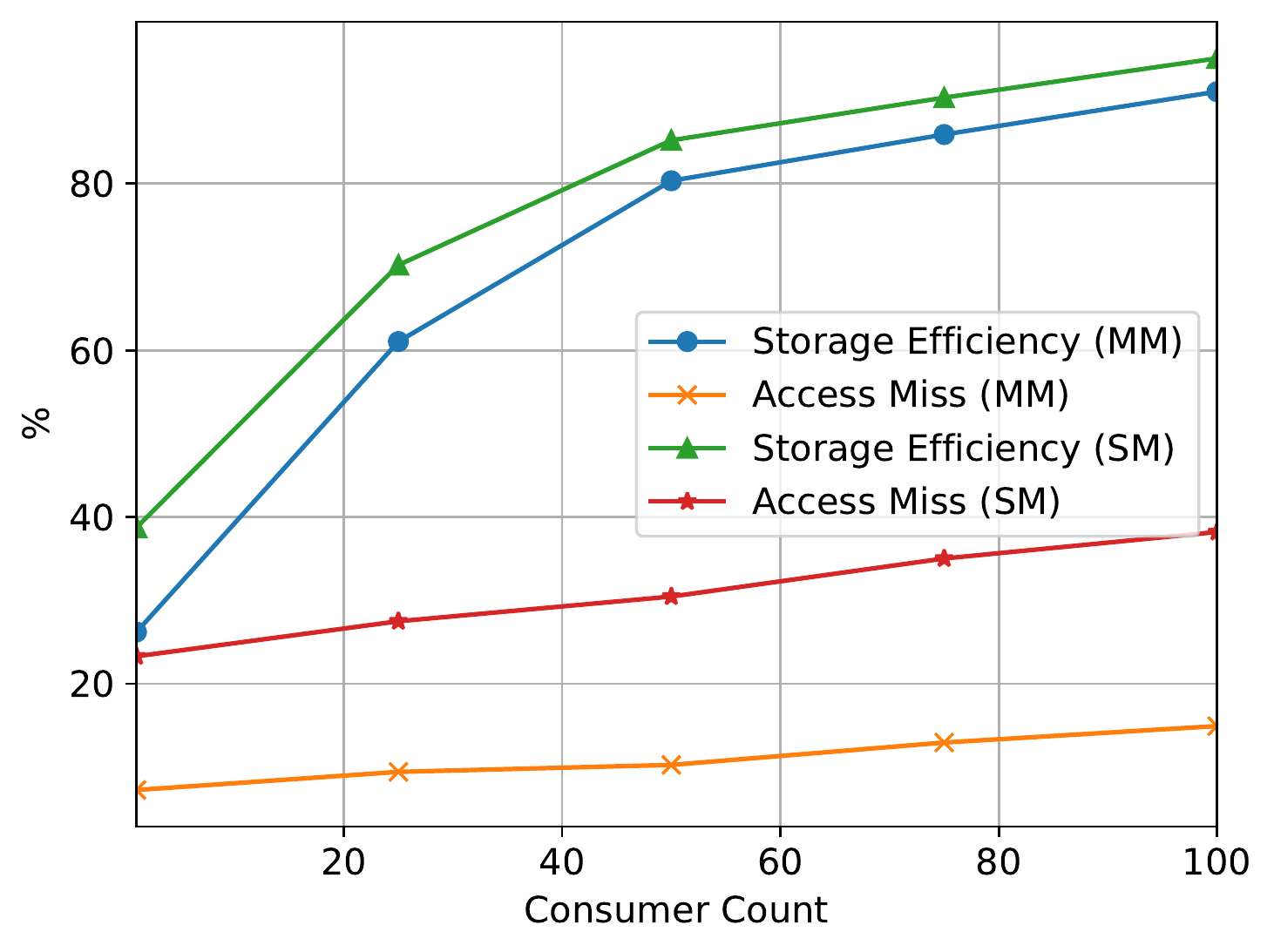}
    \caption{The storage efficiency and access misses of the framework when varying the number of consumers (Brightkite).}
    \label{figure:consumer_count_brightkite}
\end{minipage}
\end{figure*}

\begin{figure*}[!t]
\centering
\begin{minipage}{.42\textwidth}
    \centering
    \includegraphics[width=1.0\textwidth]{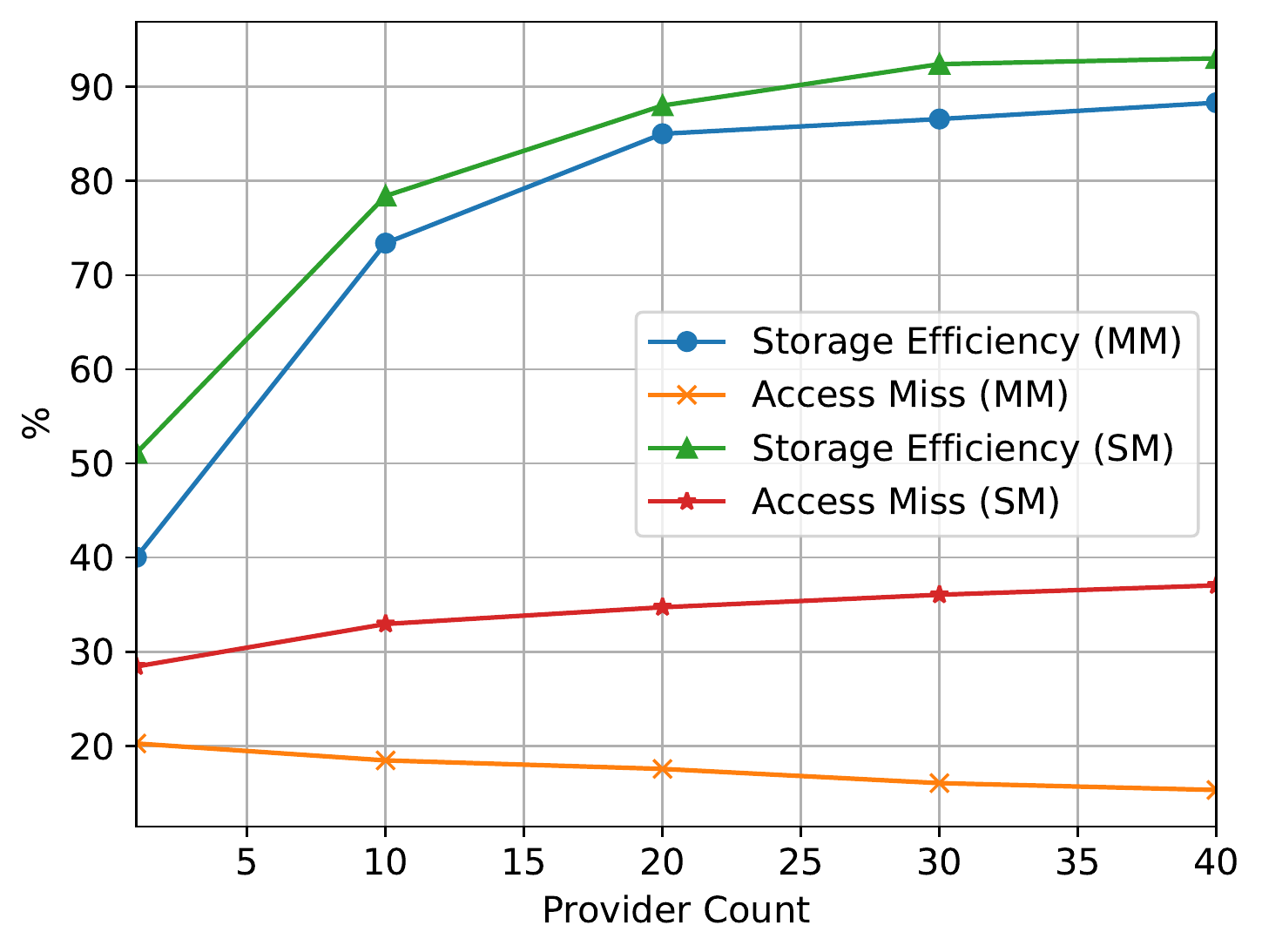}
    \caption{The storage efficiency and access misses of the framework when varying the number of providers (Gowalla).}
    \label{figure:provider_count_gowalla}
\end{minipage}%
\hspace{0.5cm}
\begin{minipage}{.42\textwidth}
    \centering
    \includegraphics[width=1.0\textwidth]{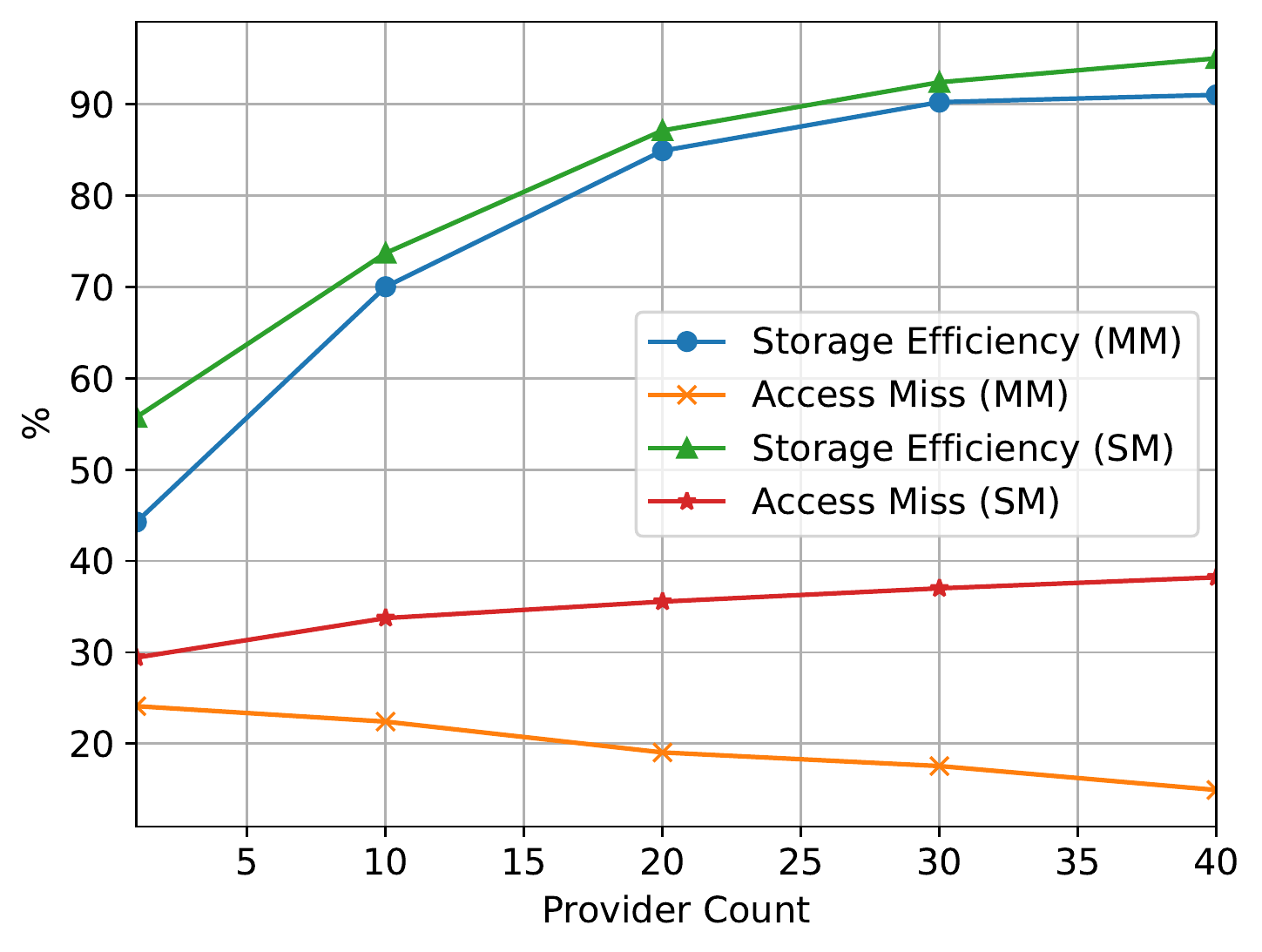}
    \caption{The storage efficiency and access misses of the framework when varying the number of providers (Brightkite).}
    \label{figure:provider_count_brightkite}
\end{minipage}
\end{figure*}

\section{Evaluation}
\label{section:evaluation}
We conduct a set of experiments to examine the efficiency of our proposed framework. We define two metrics to assess the quality of the information scoping. On one hand, the metrics ensure that information needed by a group of scoped microcells always exists in the scope. On the other hand, information that is generally not needed by the microcell group should be kept outside the scope.

\subsection{Dataset Description}
\label{section:dataset_description}
We use two real datasets in our experiments, namely, Gowalla and Brightkite \cite{cho2011friendship}. Both datasets are location-based social networks. Essentially, each dataset represents a graph, where each node in the graph is a user and an edge indicates a friendship relation between two users. Additionally, the dataset includes check-in information for every user. In that respect, each user has a list of locations at which they have checked in. Details about every the two datasets are listed in Tables \ref{table:gowalla} and \ref{table:brightkite}.

We map the two datasets to our problem as follows. Every location in the dataset is assumed to be a microcell. IoT service providers are represented by the users in the datasets. Check-in information is used to infer the movements of service providers between microcells (i.e., locations). For example, assume user $1$ in one dataset checked in to locations, $A$, $B$, and $C$. In such a case, user $1$ (i.e., service provider) is assumed to have moved between the three locations (i.e., microcells) and provided their services at them. 

\begin{table}[!t]
    \renewcommand{\arraystretch}{1.3}
    \caption{Gowalla dataset.}
    \centering
    \begin{tabular}{|c||c|}
         \hline
         Data & Count  \\ \hline
         Nodes & 196,591 \\ \hline
         Edges & 950,327 \\ \hline
         Check-ins & 6,442,890 \\ \hline
         Locations & 1,280,969 \\ \hline
    \end{tabular}
    \label{table:gowalla}
\end{table}

\begin{table}[!t]
    \renewcommand{\arraystretch}{1.3}
    \caption{Brightkite dataset.}
    \centering
    \begin{tabular}{|c||c|}
         \hline
         Data & Count  \\ \hline
         Nodes & 58,228 \\ \hline
         Edges & 214,078 \\ \hline
         Check-ins & 4,491,143 \\ \hline
         Locations & 772,966 \\ \hline
    \end{tabular}
    \label{table:brightkite}
\end{table}

\subsection{Metrics}
\label{section:metrics}
We define two metrics to assess the proposed framework: \emph{storage efficiency} and \emph{access misses}. The storage efficiency $\mathcal{SE}$ metric examines how efficient the scoped microcells at leveraging their storage space. More precisely, it is the ratio of used trust information to the unused trust information inside a scope. The metric can be obtained as follows:

\begin{equation}
    \mathcal{SE} = \frac{I_{used}}{I_{unused}}
\end{equation}
Where $I_{used}$ and $I_{unused}$ is the used and unused information record count in a scope, respectively. The access misses $\mathcal{AM}$ metric is proportional to the times trust information that does not exist in a scoped group is requested by a member of the scope. Alternatively, the access misses metric can be represented as follows:

\begin{equation}
    \mathcal{AM} = \frac{M}{A}
\end{equation}
Where $M$ is the number of misses, and $A$ is the total number of information accesses.

\begin{figure*}[!t]
\centering
\begin{minipage}{.42\textwidth}
    \centering
    \includegraphics[width=1.0\textwidth]{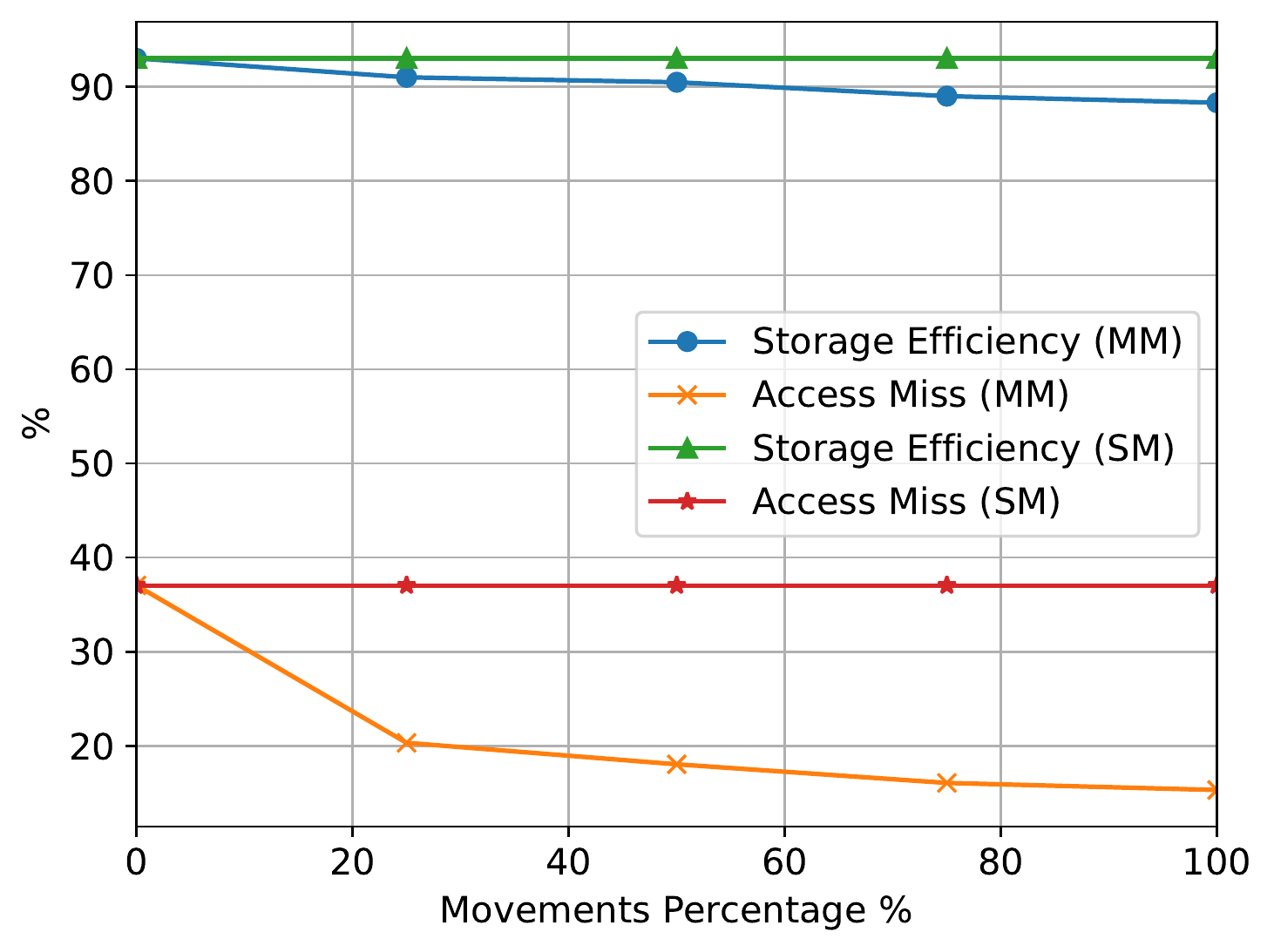}
    \caption{The storage efficiency and access misses of the framework when varying the percentage of included movements between microcells (Gowalla).}
    \label{figure:movement_count_gowalla}
\end{minipage}%
\hspace{0.5cm}
\begin{minipage}{.42\textwidth}
    \centering
    \includegraphics[width=1.0\textwidth]{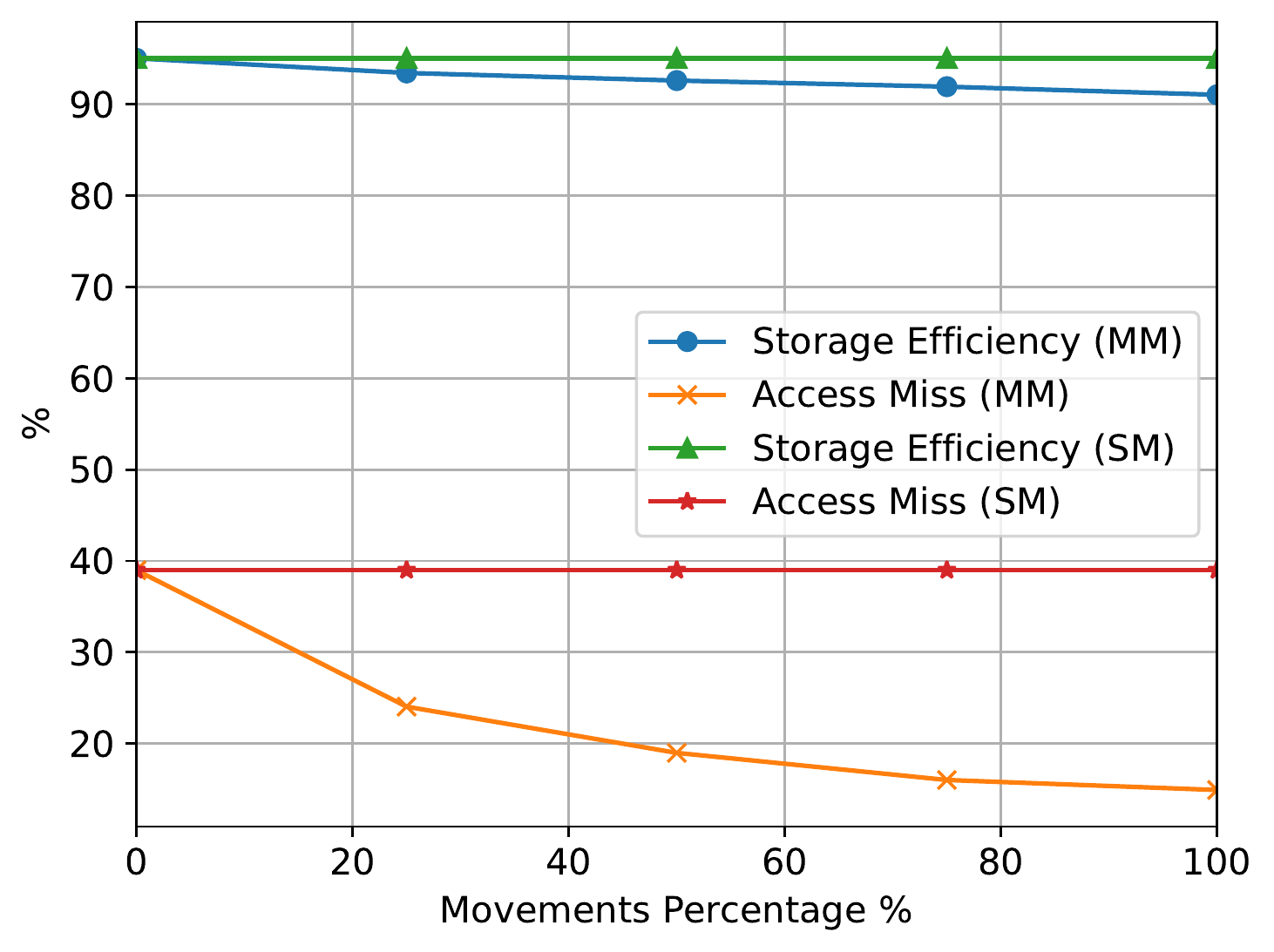}
    \caption{The storage efficiency and access misses of the framework when varying the percentage of included movements between microcells (Brightkite).}
    \label{figure:movement_count_brightkite}
\end{minipage}
\end{figure*}

\subsection{Experimental Results}
\label{section:experimental_results}
We conduct three sets of experiments to evaluate our approach. For every set, we compare our proposed framework with another variation of the framework. This variation scopes the trust information for every microcell rather than a group of microcells, thus we refer to it as Single-Microcell scoping (SM). The proposed framework scopes multiple microcells into a single scope. Therefore, we refer to it as Multi-Microcell scoping (MM). Our aim is to highlight the significance of scoping multiple microcells together rather than having a single scope for every microcell.

The first experiment examines the framework by varying the number of consumers at each microcell. Specifically, we randomly select a total of 5,000 microcells and simulate the provisioning of 40 IoT services at each one of them. The number of consumers at each microcell is changed from one consumer to 100. The experiments are conducted in both datasets. The results are shown in Fig. \ref{figure:consumer_count_gowalla} and \ref{figure:consumer_count_brightkite}. For both MM and SM scoping, the storage efficiency increases as the number of consumers increase (from 20\% to 88\%, and 26\% to 91\% for Gowalla and Brightkite, respectively in MM scoping). The rationale behind this could be that when a small set of consumers exist, only a small part of the information is accessed. Therefore, the microcells are storing information that they do not use frequently. As the number of consumers increases, the chances of accessing more trust information increases as well. Access miss also slightly increases as the number of consumers increases (from 6\% to 15\%, and 7\% to 15\%, for Gowalla and Brightkite, respectively in MM). This increase can be attributed to the increased chance of accessing information for providers that do not have any information stored in the scope. The difference between the two scoping methods (i.e., MM and SM) is high (around 22\% - 24\%). One reason may be that, with SM scoping, microcells do not share information, which leads to less information per scope. This in turn could lead to a higher probability of not finding the needed information in a microcell.

The second set of experiments evaluates the framework by varying the number of providers at each microcell. For this, we randomly select 5,000 microcells and simulate a varying number of service providers. More specifically, We start with one provider at each microcell and increase the number to up to 40 providers. For each scenario, we simulate a total of 100 consumers. The results of the experiments are shown in Fig. \ref{figure:provider_count_gowalla} and \ref{figure:provider_count_brightkite}. The storage efficiency of the framework increases as the number of providers increases in the microcell for both MM and SM scoping (40\% to 88\%, and 44\% to 91\%, for Gowalla and Brightkite, respectively in MM scoping). The reason behind this is somewhat similar to the earlier scenario; increasing the number of providers gives the microcell the chance to use more of its trust information, thus increasing the efficiency of its storage. Conversely, the access miss in MM scoping decreases as the number of providers increases (20\% to 15\%, and 24\% to 15\%, for Gowalla and Brightkite, respectively). Increasing the number of providers could decrease the probability of having a provider that does not have its information stored in the microcell's scope, thus minimizing the number of misses with respect to the total number of accesses (lower access miss). In the case of SM scoping, the access increases as the number of providers increase. As a result, the difference between SM and MM scoping increases when more providers are added (having many providers is expected in such environments). The difference between the two scoping methods reaches 22\% - 24\%. The reason behind this increase could be because of the limited trust information inside each microcell in the case of SM scoping.

The final experiment set studies the effect of the number of movements between microcells. Recall that information scoping is carried out by detecting communities in the microcell graph. The microcell graph is obtained by relying on the microcells and the movements happening between them. A provider moving between two microcells would essentially create an edge between them. In this experiment set, we change the number of movements and examine the effectiveness of the framework. More specifically, we start by removing all movements between microcells and gradually increasing the number of movements to 25\%, 50\%, 75\%, and 100\% of the original movements. For each scenario, we simulate a total of 40 providers and 100 consumers at each microcell. The results of this experiment is shown in Fig. \ref{figure:movement_count_gowalla}, and \ref{figure:movement_count_brightkite}. As shown in the figures, the storage efficiency slightly decreases as more movements are considered in MM scoping (93\% to 88\%, and 95\% to 91\%, for Gowalla and Brightkite, respectively). In the case where all movements are removed, each microcell has its own information scope. Therefore, each microcell would utilize almost all of its trust information. As we add more movements, microcells start sharing their trust information together, which increases the number of unusable trust information. Similarly, the access miss also decreases as more movements are included in MM scoping (37\% to 15\%, and 38\% to 15\%, for Gowalla and Brightkite, respectively). More movements mean a less sparse graph that leads to more information sharing between microcells, which in turn leads to fewer access misses. As for the SM scoping, the movements between microcells do not affect its performance. This is due to the fact that, unlike MM scoping, generating the scopes do not depend on such movements. 

\section{Related Work}
\label{section:related_work}
Assessing trust in crowdsourced IoT environments is fairly new. Several trust management frameworks have been proposed that aim to address trust in IoT \cite{chen2011trm, malik2009trust, saied2013trust, nitti2012subjective, wahab2020endorsement, kantarci2014mobility}. One major challenge that needs to be accounted for is storing data and information used by such frameworks. Any approach that proposes to address such a challenge should (1) be distributed to fit the dynamic nature of IoT, (2) preserve the information's integrity, and (3) available whenever needed. Several approaches and techniques were proposed to handle data storage in distributed environments. The work in \cite{adya2002farsite} proposed Farsite federated storage that runs on an inherently untrusted environment. It utilized a Byzantine-fault-tolerant protocol to achieved data integrity and cryptographic algorithms to ensure its confidentiality. Another approach proposed in \cite{zimmermann2016maintaining} a peer-to-peer protocol for sharing data based on the trustworthiness of users. Essentially, keeps track of the users' reputations, and any data coming from untrustworthy services are rejected. The work in \cite{chervyakov2019ar} proposed a data storage framework that utilizes the Redundant Residue Number System (RRNS), which leads to increased safety and reliability and faster data encryption processing. Another work proposed a system that allows organizations to store IoT data on the cloud \cite{bokefode2016developing}. Their system relies on access control policies and cryptographic concepts to ensure the security of the data. The work in \cite{wang2009secure} proposed a technique for preserving the confidentiality of outsourced data. The approach offers flexible cryptography-based access control by using multiple keys for encrypting different data blocks.

The aforementioned approaches have several properties that make them unsuitable to store trust information in IoT environments. Byzantine-fault-tolerant-based approaches require a centralized authority that manages the members of the network. However, in IoT environments managing IoT devices is impractical because of its dynamism. Other approaches focus mainly on confidentiality. While confidentiality is critical in certain scenarios, it is less in the case of trust information. Therefore, a new storage framework is needed that is specifically tailored for storing trust information in IoT environments.

\section{Conclusion}
\label{section:conclusion}
We proposed a framework for storing trust information in crowdsourced IoT environments. The framework preserves the integrity of the data by leveraging the blockchain as the primary storage medium. IoT devices and edge servers at each microcells act as the nodes of the blockchain for validating new blocks. Trust information stored by the framework is scoped. In that respect, every group of microcells that might share the same trust information would have its own blockchain network. Therefore, the blockchain would store trust information that is truly needed by the microcell, and thus, storage space would be preserved. Our conducted experiments show the effectiveness of the proposed framework. In our future work, we will extend the framework to handle the dynamic aspects of the microcells. More specifically, the framework will be developed to adapt to changes in movement patterns and handle obsolete trust information.

\section{Acknowledgements}
This research was partly made possible by DP160103595 and LE180100158 grants from the Australian Research Council. The statements made herein are solely the responsibility of the authors.

\bibliographystyle{IEEEtran}
\bibliography{IEEEabrv, ./ref.bib}

\end{document}